\documentclass[conference]{IEEEtran}

\ifCLASSINFOpdf
\else
\fi
\usepackage[T1]{fontenc}
\usepackage{amsmath}
\usepackage{glossaries}
\usepackage{graphicx}
\usepackage[capitalise,noabbrev]{cleveref}
\usepackage{float}
\usepackage{url}
\usepackage{listings}
\usepackage{comment}
\usepackage{subcaption}
\usepackage[numbers,sort&compress]{natbib}
\usepackage{multirow}
\usepackage[table,xcdraw]{xcolor}
\usepackage{booktabs}
\usepackage{array}
\usepackage{tabularray}
\usepackage{threeparttable}
\UseTblrLibrary{booktabs}
\usepackage{siunitx}

\usepackage{amssymb}

\providecommand{\ie}{\emph{i.e.,} }
\providecommand{\eg}{\emph{e.g.,} }

\newcolumntype{Y}{>{\raggedright\arraybackslash}X}

\begin{document}

\title{\Large \bf Is That Really My X-Ray?\\ 
Measuring Internet-Exposed DICOM Services in the Presence of Deception}


\author{\IEEEauthorblockN{Ricardo Yaben, Karina Elzer, Alberto Maria Mongardini, and Emmanouil Vasilomanolakis}
\IEEEauthorblockA{Department of Applied Mathematics and Computer Science\\
Technical University of Denmark, Kgs. Lyngby, Denmark\\
Email: \{rmyl, kaelz, among, emmva\}@dtu.dk}}


%



\maketitle

\begin{abstract}
DICOM is the dominant protocol for exchanging medical images, yet many Internet-facing deployments lack basic security controls, exposing sensitive patient data to unauthorized access.
Accurately measuring this exposure is complicated by honeypots, network telescopes, and other measurement artifacts that inflate published estimates.
This paper presents a noise-aware study of Internet-facing DICOM services, combining active IPv4-wide scanning with passive honeypot deployments. We scan common DICOM ports using an ethics-constrained probe limited to association negotiation and C-ECHO. Then, we introduce a reproducible false-positive filtering method that identifies honeypots, telescopes, malformed responders, and other deception artifacts, reducing apparent exposure by 39\%.
After filtering, we identify 3,979 vulnerable DICOM deployments, all of which lack encryption and accept unauthenticated connections; 1,782 run outdated or deprecated software, and 1,551 carry known remotely exploitable vulnerabilities. 
Follow-up scans reveal that approximately 50\% of exposed services show no evidence of maintenance over 5 months of observation, and that responsible disclosure led to only modest, short-term remediation.
On the passive measurement, we deploy Dicompot and find that its raw session logs overstate activity by up to 83\% due to generic TCP noise being incorrectly logged as DICOM  sessions. After filtering, we observe a reconnaissance gap: Most actors issuing C-ECHO probes never escalate to data exfiltration, consistent with sophisticated actors fingerprinting the honeypot and disengaging before proceeding. 
Our results show that DICOM exposure measurements can be significantly distorted by deception and logging artifacts, but once corrected, reveal widespread risks to patient privacy and healthcare security that existing deception systems are insufficient to capture. 

\end{abstract}


%
\IEEEpeerreviewmaketitle

\section{Introduction}
The medical sector has become an increasingly attractive target for cyberattacks that threaten patient data integrity, disrupt healthcare delivery, and compromise clinical operations. In 2023, the European Commission reported 309 cybersecurity incidents affecting the sector~\cite{eu_cyber_report}, motivating a dedicated healthcare cybersecurity action plan. The trend continued in 2024: Ransomware attacks against Ascension Health~\cite{ascension2024blackbasta} and Change Healthcare~\cite{aha2024changehealthcare} forced hospitals to cancel procedures, divert ambulances, and return to manual charting, while similar disruptions affected Lurie Children's Hospital~\cite{lurie2024recovery} and the OneBlood distribution center~\cite{aha2024oneblood}. Beyond operational disruption, these incidents often result in the theft of sensitive medical records---including prescriptions, clinical images, and diagnostic information---which makes healthcare data a persistent target for financially motivated attackers~\cite{healthcare8020133}.


Among the protocols supporting modern healthcare infrastructure, \gls*{dicom} is both critical and frequently exposed~\cite{censys_2024_ioht}. Since its introduction in 1993, \gls*{dicom} has become the international standard for transmitting, storing, and displaying medical images, including \gls*{mri}, \gls*{ct}, and X-ray scans~\cite{DICOM}. Yet, despite handling sensitive patient data, many \gls*{dicom} security mechanisms remain optional or are not consistently enforced~\cite{dicom_sec}. The consequences are visible at Internet scale: CybelAngel~\cite{fullblood} reported millions of unsecured medical records across more than 2,140 exposed services, and Yazdanmehr~\cite{apilite2023} identified 3,806 publicly accessible \gls*{dicom} servers, fewer than 1\% of which employed effective authorization.

In this work, we focus on Internet-facing \gls*{dicom}-speaking services as a representative and high-impact target for healthcare security measurement. While prior studies have shown that exposed \gls*{dicom} deployments exist at scale, measuring their real prevalence and security posture remains challenging: Internet-wide scans encounter not only production systems, but also honeypots, network telescopes, malformed responders, and other artifacts that can inflate exposure estimates. We therefore provide a comprehensive, noise-aware analysis of Internet-facing \gls*{dicom} services, designed to separate real deployments from measurement artifacts before assessing their security implications. To minimize risk to patients and operators, our active measurements follow an ethics-constrained methodology: probes are limited to association negotiation and C-ECHO, avoid authentication attempts and data-access commands, honor opt-out requests, and never retrieve or inspect medical records.


Our study combines active and passive measurements to characterize exposed \gls*{dicom} services and the artifacts that affect their interpretation. In the active component, we conduct an Internet-wide survey of common \gls*{dicom} ports to evaluate deployment weaknesses related to encryption, authentication, and access control. 
We further fingerprint exposed implementations to identify products, versions, and potentially vulnerable software, finding 1,782 services running outdated or deprecated software, most affected by known vulnerabilities. 
In the passive component, we deploy Dicompot~\cite{nsmfoo_dicompot}, a \gls*{dicom} honeypot emulating a vulnerable RadiAnt workstation~\cite{radiant}, from both a cloud provider and our institution's network to observe unsolicited Internet traffic and assess how passive logs can be distorted by non-\gls*{dicom} background noise.

The artifacts from this study can be found in our open science repository\footnote{\url{https://anonymous.4open.science/r/DICOM_Internet_Measurement_Paper-8A55}}, along with datasets, probe, summaries, and analysis pipeline. 
Our contributions are as follows:

\begin{itemize}
    \item \textbf{Comprehensive Internet-Wide Scan.}
    We conduct an Internet-wide scan of \gls*{dicom} services across common deployment ports to measure real exposure at scale. 
    To improve the accuracy of this measurement, we fingerprint sources of noise and deception, including honeypots, network telescopes, malformed responders, and other artifacts that distort exposure estimates. 
    Our filtering method reduces false positives (Noise) by 39\%, providing a more accurate picture of Internet-facing \gls*{dicom} deployments. We further repeat our measurements before and after responsible disclosure, observing a modest but positive shift in host states: fewer services continued to accept C-ECHO requests, more became unreachable, and several deployments showed changed behavior or apparent updates. 
    Unlike prior work that retrieved or enumerated patient records, our methodology is ethics-constrained. 
    We limit active probes to association negotiation and C-ECHO, avoid authentication attempts and data-access commands, honor opt-out requests, and never retrieve or inspect medical data.
    

    
    \item \textbf{Systematic Security Analysis.} 
    We identify widespread security weaknesses across 3,979 exposed services, all of which lack encryption and accept unauthenticated interaction; 1,782 of these run on outdated or deprecated software, and 1,551 contain known vulnerabilities. 
    These issues are not limited to a single product, region, or deployment type, highlighting the need for stronger default security, better maintenance practices, and clearer operational guidance for \gls*{dicom} systems handling clinical data.


    \item \textbf{Passive Measurement \& Deception Analysis.}
    We conduct a longitudinal 324-day study of Internet traffic targeting \gls*{dicom} services by deploying Dicompot across two vantage points. Our findings reveal critical limitations in Dicompot, including low logging fidelity and high detectability, that fundamentally distort threat assessments. Specifically, we identify a reconnaissance gap in which sophisticated actors are likely to fingerprint the honeypot during the initial C-ECHO phase, resulting in a disproportionate ratio of basic connectivity probes to actual data-harvesting attempts.

    
\end{itemize}

\section{Background}
\label{sec:background}

\gls*{dicom}~\cite{DICOM} is an international standard for transmitting, storing, retrieving, and displaying medical imaging data. First published in 1993, it remains the dominant protocol for exchanging radiological images (MRI, CT, X-ray) and associated clinical metadata across hospital networks, between medical institutions, and increasingly over the Internet.

\subsection{\gls*{dicom} Files and Metadata}
\label{subsec:dicom_metadata}

A \gls*{dicom} file (typically with a \texttt{.dcm} extension) combines medical image pixel data with extensive metadata describing the patient, clinical context, and acquisition parameters. This metadata is organized as a series of data elements (tags) that enable medical images to be consistently interpreted across different systems and institutions.

Critically for security, \gls*{dicom} files contain highly sensitive patient information, including patient names, dates of birth, medical record numbers, referring physician identities, hospital identifiers, and detailed clinical notes. The metadata also includes technical parameters such as imaging modality (CT, MRI, X-ray), acquisition timestamps, and device serial numbers. When \gls*{dicom} services are exposed to the Internet without proper access controls, this sensitive data becomes accessible to unauthorized parties.

Specifically, files are organized hierarchically: a \textit{patient} may have multiple \textit{studies} (medical examinations), each study contains one or more \textit{series} (sets of related images), and each series comprises individual \textit{images} or composite objects. This hierarchy is maintained through unique identifiers assigned at each level, enabling efficient querying and retrieval across distributed archives.

\subsection{\gls*{dicom} Network Protocol}
\label{subsec:dicom_proto}
As registered with \gls*{iana}, \gls*{dicom} primarily uses TCP ports 104 and 11112 for unencrypted communications~\cite{DICOMUp9}. Encrypted channels are supported via ports 2761 and 2762 for the Integrated Secure Communication Layer (ISCL) and TLS, respectively~\cite{BSecureT43}, though these are rarely deployed in practice. Additionally, some \gls*{dicom} server implementations use non-standard default ports; notably, Orthanc, a widely-used open-source PACS server, defaults to port 4242~\cite{Configur14:online}.

Before executing any data operations, \gls*{dicom} clients must establish an association with the server through an A-ASSOCIATE handshake. 
The structure of an association request contains several key fields relevant to our analysis:

\begin{itemize}
    \item \textbf{Calling AE Title}: Identifies the requesting client, typically containing the name of the \gls*{dicom} viewer software or medical device.
    \item \textbf{Called AE Title}: Specifies the target server application entity that the client wishes to connect to.
    \item \textbf{User Information Item}: Contains optional sub-items for authentication and implementation details, including: the \textit{Implementation Class UID}, a unique identifier revealing the software vendor and product; the \textit{Implementation Version Name}, that indicates the specific software version, enabling identification of deprecated or vulnerable implementations; and the \textit{User Identity Negotiation}, which provides authentication credentials in various formats.
\end{itemize}


In particular, the AE Titles serve as a compatibility and routing mechanism, not as an authentication mechanism. While servers may reject associations based on unrecognized AE Titles, this provides minimal security, since titles are transmitted in plaintext and can be easily spoofed. The critical security weakness lies in the User Identity negotiation: the \gls*{dicom} standard defines this sub-item as \textit{optional} and provides no mandatory authentication requirements. The standard suggests supporting usernames, username/password pairs, Kerberos service tickets, or SAML assertions~\cite{DICOM}, but explicitly states that servers may ignore the User Identity sub-item entirely.
This optionality creates a systemic vulnerability. While the standard recommends TLS encryption for credential protection and discourages plaintext passwords, it provides no enforcement mechanism. In practice, many \gls*{dicom} implementations ship with authentication disabled by default, and deployers often fail to enable it, leading to the widespread exposure we document in \cref{sec:exposure}.

If a server accepts the association request, it responds with A-ASSOCIATE-AC (acknowledgment), potentially negotiating security parameters. If rejected, it returns A-ASSOCIATE-RJ with a reason code. Once associated, clients can execute \gls*{dicom} operations without further authentication challenges in most implementations.

\subsubsection{\gls*{dicom} Operations}

After establishing an association, clients interact with \gls*{dicom} servers through \gls*{dicom} Message Service Element (DIMSE) commands. The operations relevant to our study include:

\begin{itemize}
    \item \textbf{C-ECHO}: Verifies network connectivity and server responsiveness. This minimal operation sends and receives a single packet with negligible overhead.
    \item \textbf{C-FIND}: Queries the \gls*{dicom} database using attribute filters (\eg patient name, study date). The server returns a list of matching entries with requested attributes.
     \item \textbf{C-GET}: Retrieves matching \gls*{dicom} files directly over the association connection. The server streams files to the client based on query results. This enables complete data exfiltration when combined with broad C-FIND queries.
     \item \textbf{C-MOVE}: Transfers matching \gls*{dicom} files to a specified third-party destination. Unlike C-GET, the files are not sent to the requesting client but to an arbitrary network address specified in the request. This can be abused for data exfiltration to attacker-controlled servers.
     \item \textbf{C-STORE}: Transmits and stores \gls*{dicom} files on the server. When authorization is absent, attackers can inject falsified medical records or maliciously crafted files.
\end{itemize}

\begin{figure}
    \centering
    \includegraphics[width=.8\columnwidth]{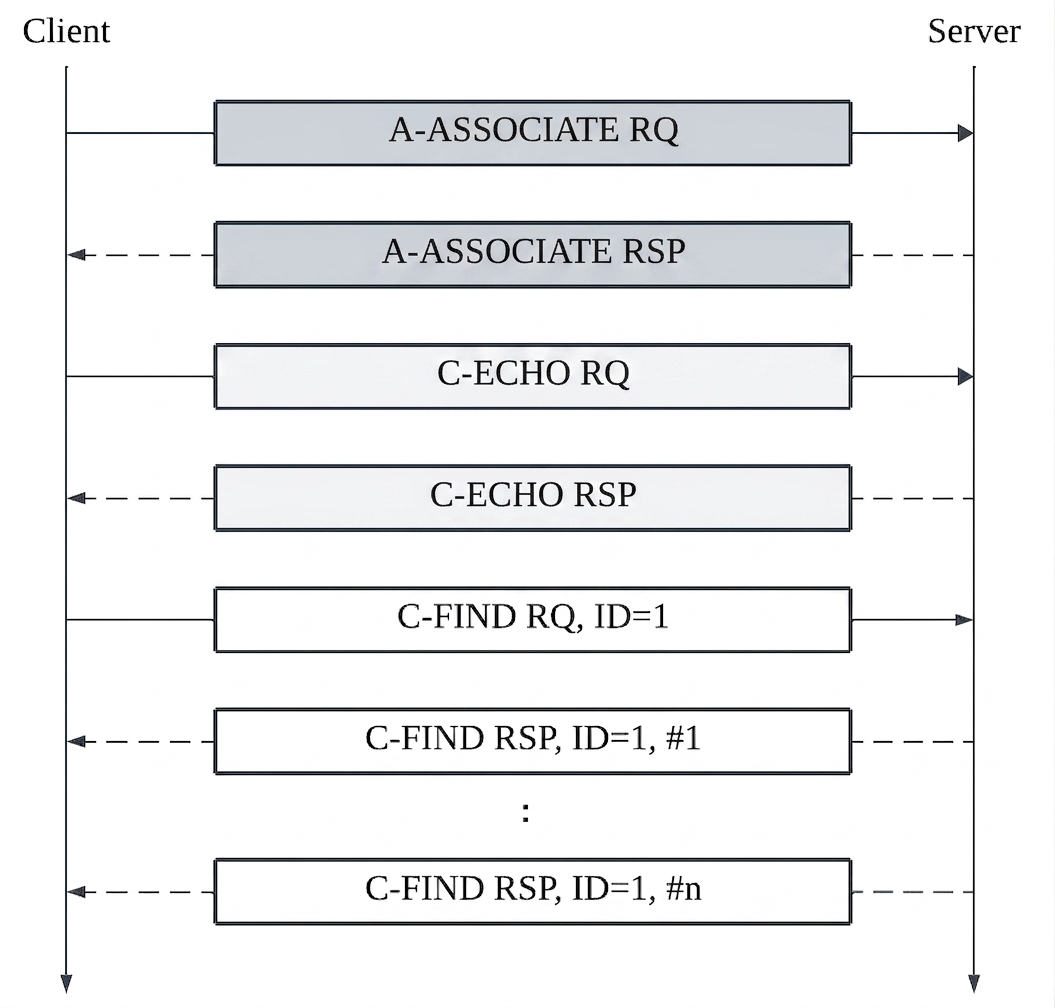}
    \caption{Example of \gls*{dicom} communication between a client and a server. The sequence begins with an association request (A-ASSOCIATE), followed by a C-ECHO command to verify the server's health, and a subsequent query to retrieve studies (C-FIND).}
    \label{fig:dicom-uml}
\end{figure}

\cref{fig:dicom-uml} illustrates a typical interaction sequence: the client establishes an association with A-ASSOCIATE, verifies server health with C-ECHO, then issues a C-FIND query to search for studies. This pattern mirrors both legitimate clinical workflows and the reconnaissance-to-exploitation progression we observe in attacker behavior (\cref{sec:active}).

Critically, the \gls*{dicom} standard provides no mandatory authorization framework for these operations. While implementations may restrict which operations authenticated users can perform, the standard does not require differentiation between C-ECHO (minimal impact) and C-GET (complete data exfiltration). This design decision, combined with optional authentication, creates the vulnerability landscape we measure: Services that accept associations from arbitrary Internet clients typically permit unrestricted execution of all DIMSE commands, enabling the attack vectors described in our threat model (\cref{sec:threat-model}).
\section{Related Work}
\label{sec:related}
The security of the \gls*{dicom} protocol and the broader landscape of Internet-facing medical services have attracted growing attention. Yet, prior work remains fragmented across three largely disconnected threads: protocol-level vulnerability analysis, Internet-wide active measurements, and passive monitoring via deception systems.

\subsection{Protocol Vulnerabilities}

\label{sec:rel-protocol-vuln}
While the \gls*{dicom} protocol has facilitated medical imaging for decades, it harbors vulnerabilities that expose healthcare systems to many cyber threats~\cite{Pianykh2008}. 
\citet{Milena2021} investigated hidden transmission channels within \gls*{dicom}, while \citet{Desjardins2020} described attack classes ranging from unauthorized data access to malicious injection and called for coordinated stakeholder responses.
\citet{Wang2021} applied fuzzing to uncover implementation-level vulnerabilities in \gls*{dicom} software stacks, highlighting the need for continuous security auditing.
These works establish the theoretical attack surface; our study operationalizes it by measuring how that surface manifests across Internet-facing deployments in the wild.

\subsection{Active Measurements}
\label{sec:rel-active}

Empirical studies of \gls*{dicom} exposure at Internet scale are scarce,  and as summarized in \cref{tab:related-active}, they share several methodological limitations.
\citet{StitesPianykh2016} conducted one of the first large-scale analyses of Internet-exposed radiology servers, revealing widespread unprotected deployments and arguing for the adoption of international security standards. 
The authors in~\cite{CorderoBarria,CorderoVidalBarriaHuidobro} conducted targeted assessments of Latin American \gls*{dicom} nodes and identified misconfigured deployments in PACS. 
More recently,~\citet{apilite2023} performed a six-month Internet-wide scan from multiple vantage points, identifying 3,806 publicly accessible \gls*{dicom} servers of which fewer than 1\% employed effective authorization mechanisms. Their methodology explicitly filters out honeypots and unrelated services; however, no details are provided on how this filtering is performed, making their results difficult to reproduce or validate. Moreover, to quantify data exposure, they enumerate and retrieve patient records from vulnerable servers, a practice that raises serious ethical concerns and that we explicitly avoid in our study (\cref{sec:ethics}).
While these studies collectively demonstrate that the problem exists, they present critical limitations. First, they typically rely on a single-snapshot approach, failing to capture the longitudinal evolution of the \gls*{dicom} landscape. To the best of our knowledge, our study is the first to employ multiple consecutive scans over time, specifically designed to evaluate changes in the global exposure surface following a disclosure campaign. Second, \cite{CorderoBarria,CorderoVidalBarriaHuidobro} are limited to Chile, and with the exception of~\cite{apilite2023}, which acknowledges honeypots and attempts to filter them out but does not disclose the filtering methodology, none of these studies accounts for the distortion that deception systems introduce into exposure estimates.

\begin{table}
\centering
\caption{Comparison of Internet-wide \gls*{dicom} exposure studies. We mark a study as intrusive if it retrieves patient records and longitudinal if it performs follow-up scans.}
\label{tab:related-active}
\begin{tabular}{@{}lrllcc@{}}
\toprule
\textbf{Study} & \textbf{Servers} & \textbf{Scope} & \textbf{Noise Filter.} & \textbf{Intrusive} & \textbf{Longit.} \\
\midrule
\cite{StitesPianykh2016}               & 2,774 & World & None & $\times$ & $\times$ \\
\cite{CorderoBarria}                   & 45    & Chile     & None & $\times$ & $\times$ \\
\cite{CorderoVidalBarriaHuidobro}      & 14    & Chile     & None & $\times$ & $\times$ \\
\cite{apilite2023}                     & 3,806 & World & Undisclosed & \checkmark & $\times$ \\
\textbf{This work}                     & \textbf{3,979}   & \textbf{World} & \textbf{Reproducible} & $\times$ & \checkmark \\
\bottomrule
\end{tabular}
\end{table}


A substantial body of work has studied how honeypots can be detected and 
fingerprinted. The authors of~\cite{holz2005detecting,papazis2019detecting} characterized 
attacker-side fingerprinting techniques, while~\citet{vetterl2018bitter} 
demonstrated how protocol-level inconsistencies reliably expose low- and 
medium-interaction honeypots at Internet scale.~\citet{morishita2019detect} 
discovered thousands of poorly configured honeypots. Huang et al.~\cite{huang2021honeypot} proposed machine-learning-based detection, 
and~\citet{srinivasa2023multistage} introduced a comprehensive fingerprinting 
framework identifying vulnerabilities across thousands of systems.
Complementary work has focused on making honeypots harder to fingerprint. 
\citet{provos2003honeyd} introduced the Honeyd framework, which emulates 
TCP/IP behaviors to thwart network fingerprinting, while~\citet{mohammadzadeh2013evaluation} 
highlighted the advantages of dynamic honeypots, which adjust their 
configurations to avoid detection.
None of these works, however, considers the \gls*{dicom} protocol, leaving the detectability of medical honeypots an open problem that we address in \cref{sec:exposure}.

\subsection{Passive Measurements \& Deception Systems}
\label{par:rel-passive}

Deception systems (\eg honeypots, tarpits, and telescopes) are widely used to monitor attack campaigns and detect novel exploitation patterns.
Their value in healthcare environments is recognized but understudied.
\citet{ihanus2020modelling} proposed a conceptual framework for deploying honeypots as proxies for medical devices, emphasizing their ability to detect cyber threats with a low false-positive rate.
To capture real-world botnet activity,~\citet{WANG2022108212} operationalized a related idea at scale by deploying in 22 different countries 462 honeypots that mimicked smart devices on the Internet of Medical Things (IoMT).
Within the \gls*{dicom} ecosystem, Dicompot~\cite{nsmfoo_dicompot} is the primary honeypot in active use and is distributed as part of the T-Pot framework~\cite{Deutsche_Telekom_Security_GmbH_and_Ochse_T-Pot_24_04_1_2024}. 
Dicompot~\cite{nsmfoo_dicompot} is a low-interaction honeypot specifically designed to simulate the \gls*{dicom} protocol by implementing a subset of DIMSE-C commands (\eg C-FIND, C-MOVE, C-GET) but does not populate stored files with realistic metadata and provides no simulation of authentication or encryption---all characteristics that make it detectable at the protocol level, as we show in \cref{sec:exposure}. 

The most directly comparable prior deployment study is~\cite{10115073}, who used Dicompot to observe attacks on \gls*{dicom} over one month in 2022 using a single vantage point within a university network. While the authors recorded 61 sessions on \gls*{dicom}, their methodology was part of a broader study of automated traffic across 14 protocols, rather than a dedicated analysis of \gls*{dicom}, which limits the conclusions that can be drawn about specific \gls*{dicom} attacker behavior and intent. 
First, their analysis relied on raw session counts without applying noise-filtering heuristics; as a result, their findings do not distinguish between malicious intent and the background noise of internet scanners or research probes. 
Second, they mention the presence of C-FIND commands without quantifying them and omitting to analyze the query payloads, thereby preventing a distinction between targeted data harvesting (\eg wildcard searches) and simple protocol discovery. 

Our work addresses these limitations directly.
On the active measurement side, we introduce a noise-filtering method that detects Dicompot instances, telescopes, and anomalous \gls*{dicom} behaviors. Furthermore, our study is the first to employ consecutive Internet-wide scans to monitor the evolution of the \gls*{dicom} landscape. This longitudinal approach includes a post-disclosure scan specifically designed to quantify the effectiveness of our remediation outreach.
On the passive measurement side, we utilize a long-term Dicompot deployment to characterize the nature of real-world \gls*{dicom} interactions, distinguishing between network noise and targeted data-harvesting attempts.
Together, these contributions provide the first comprehensive, noise-aware view of \gls*{dicom} exposure and attacker behavior on the Internet.
\section{Methodology}
\label{sec:methods}


In January 2026, we conducted an IPv4 Internet-wide scanning campaign of \gls*{dicom}-speaking services facing the Internet ($S1$).
This campaign employed a surveying strategy, scanning the entire address space three consecutive times.
While still a cross-sectional study, surveying the internet multiple times has been shown to improve measurement accuracy~\cite{10.1145/3718958.3754344}. This approach mitigates issues stemming from the dynamic nature of the network, such as churn, jitter, and fluctuating availability.
This is particularly relevant for experiments where time variations may introduce additional measurement biases.
In the case of \gls*{dicom}, we expect workstations to have alternating schedules (\eg \gls*{dicom} viewers located in daily clinics and research institutions), as they are commonly hosted on regular computers and not on servers.
We conducted two additional scans of the vulnerable services we identified, once in April ($S2$) before notifying the affected stakeholders, and again in May ($S3$) after communicating our findings.

\paragraph{Ports}

Based on IANA's registered ports and findings from previous authors~\cite{censys_2024_ioht, aplite2023}, our scans interrogate five default ports commonly used for \gls*{dicom} communications: 104 and 11112 used for standard communications; 2761 and 2762 designated for encrypted channels; and 4242 specific to Orthanc services.
Orthanc services have been widely reported to be one of the most commonly seen \gls*{dicom} services facing the Internet.
Note that this study does not aspire to provide the completeness of an Internet Census of exposed \gls*{dicom} services, leaving the task of identifying exposed services in non-standard ports for further research.
Moreover, our experiments do not consider web portals or other services giving access to medical records.
Thus, the results provided in this paper should be interpreted as lower bounds rather than in absolutes.

\paragraph{Vantage Point}

Scans were carried out from a single vantage point located within our institution's network, which hosts a website with details regarding our measurements and instructions on how to opt out of our studies.
We honor current and past requests using a blacklist of addresses that do not wish to be included or scanned.
This vantage point has been used in multiple similar experiments, which may reduce its visibility.
We acknowledge that using a single vantage point further degrades the coverage of our study~\cite{wan_origin_2020}.

\paragraph{Instrumentation}
Furthermore, we used the ZMap toolkit to carry out each iteration of scanning, first sweeping the targeted ports with TCP SYN/ACK stateless probes with ZMap, and followed by stateful \gls*{dicom} probes with ZGrab2.
The ZMap toolkit implements several mitigation measures to reduce the impact of our scans in the target networks (\eg address randomization, hardcoded identifiers, and timeout delays between probes toward each target).
Our configuration was limited to setting a maximum connection timeout period of 30 seconds (instead of the default 10 seconds) per address in an attempt to mitigate further representation biases from slower services and network issues affecting communications.  

\paragraph{\gls*{dicom} Probe}

This probe was designed to provide limited interaction with \gls*{dicom} services, reducing communications to an association handshake (A-ASSOCIATE) and an echo request (C-ECHO).
Further, the probe includes several identifiers to disclose the intention and the origin of the probe, namely: the probe uses custom (i)~\texttt{CallingAETitle} with the name of our institution, (ii)~Implementation UID \texttt{1.2.3.4.5}, and (iii)~Implementation Version showing \texttt{ZGRAB2} instead of a known \gls*{dicom} software.

The absence of authentication parameters and the association release request (A-RELEASE) is intentional.
Authentication attempts using arbitrary credentials could be considered brute-force attacks. 
The association release is omitted because it is not required to terminate connections.
Otherwise, the probe uses a \texttt{CalledAETitle} iterator to fulfill validation requirements in certain services, at the cost of one additional request per title.
While our scanning campaign iterates over up to five titles when the association evaluation of AE titles fails, approximately 81\% of exposed services required one iteration.

\paragraph{Exposure \& Vulnerability Taxonomy}

Based on the interactions facilitated by our \gls*{dicom} probe, we establish a strict taxonomy to classify the security posture of discovered services. This classification bridges our scanning methodology with the resulting vulnerability analysis. As summarized in \cref{table:taxonomy}, each step of our probe's interaction tests a specific security boundary: discovery, authentication, and authorization.

\begin{table}
    \centering
    \caption{\gls*{dicom} Exposure Taxonomy. We map specific probe interactions to the evaluated security boundaries.}
    \label{table:taxonomy}
    \resizebox{\columnwidth}{!}{%
    \begin{tabular}{cll}
    \toprule
    \textbf{Level} & \textbf{Security Boundary} & \textbf{Scanner Action \& Expected Response} \\ 
    \midrule
    \textbf{1} & Protocol Exposure & A-ASSOCIATE $\rightarrow$ AC or RJ \\
    \textbf{2} & Missing Authentication & A-ASSOCIATE (No Auth) $\rightarrow$ AC \\
    \textbf{3} & Authorization Failure & C-ECHO $\rightarrow$ Success \\ 
    \textbf{4} & Data Accessibility & \textit{Not Measured (Ethical Constraints)} \\
    \bottomrule
    \end{tabular}%
    }
\end{table}

We define the boundaries as follows. \textit{Exposure} (Level 1) occurs when a service responds with a valid A-ASSOCIATE acknowledgment or rejection, revealing its \texttt{Implementation Class UID} and version. At this level, we identify missing encryption if the service operates on standard unencrypted ports and known vulnerabilities by cross-referencing the UID with CVE databases. \textit{Missing Authentication} (Level 2) is confirmed when the service accepts our association request despite the intentional omission of the User Identity sub-item. \textit{Authorization Failure} (Level 3) is reached when the service processes our subsequent C-ECHO command. While C-ECHO is non-intrusive, a service accepting it from an unauthenticated client serves as a conservative proxy indicator for broader authorization failures. \textit{Data Accessibility} (Level 4) involves querying or retrieving actual medical data. To adhere strictly to ethical scanning guidelines and avoid accessing sensitive patient data, our methodology intentionally stops at Level 3. 

%

\paragraph{Third-Party Data Sources}

We enhance our post-processed scanning results (\ie identified exposed \gls*{dicom} services) with information from IPinfo~\cite{ipinfo}.
This service provides \gls*{as} intelligence, DNS records, and WHOIS metadata, valuable to notify network operators and distinguish clinics, hospitals, and research institutions from other hosts.
In particular, we use their \textit{IP to Company} database to match results from identified exposed \gls*{dicom} services.


\section{Threat model}
\label{sec:threat-model}

In this work, we adopt a threat model focused on what attackers can observe and exploit through Internet-wide scanning of \gls*{dicom} services. 
We do not assume the attacker has prior access to compromised credentials, specific vendor knowledge, or a local network presence. Instead, our model aligns with the visibility established by our scanning methodology (\cref{sec:methods}): If our benign scanner can identify, associate with, and issue commands to a \gls*{dicom} service, malicious actors can do the same. 

Our model builds on the vulnerability taxonomy proposed by~\citet{StitesPianykh2016} and extends the threat framework introduced by~\citet{Eichelberg2020} to evaluate attack vectors enabled by Internet exposure.

\subsection{Exploitable Attack Vectors}
Based on the exposure levels defined in our taxonomy (\cref{table:taxonomy}), services that reach Level 3 (Authorization Failure) are presumed susceptible to Level 4 (Data Accessibility) exploitation. This exposure enables three primary attack vectors, adapted from~\cite{Eichelberg2020} for Internet-facing deployments: (i) data exfiltration of medical images and associated metadata (\eg patient identifiers, clinical notes, diagnostic information) through C-FIND/C-GET/C-MOVE commands; (ii) data manipulation with injection of falsified medical records (C-STORE) or tampering with existing records through deletion and replacement facilitated by the \gls*{dicom} file format's flexible preamble and image encoding that provide multiple injection points for malicious payloads~\cite{75Securi6:online}; (iii) service disruption through Denial-of-service attacks through resource exhaustion (flooding with C-STORE requests), protocol abuse (malformed associations), or ransomware deployment following initial access.

\subsection{Regulatory Compliance Dimension}
\label{sec:compliance}

Beyond technical security measures, we evaluate \gls*{dicom} deployments against regulatory frameworks governing medical devices and health information systems. 

\paragraph{Medical Device Certification}

In most regions, medical imaging software requires regulatory approval before clinical deployment.
Examples include the \gls*{ce} marking, a manufacturer's declaration of compliance with EU safety standards~\cite{Implemen74:online}, and \gls*{fda} clearance, the formal validation of a device's safety and effectiveness in the United States~\cite{FDA}.
These frameworks mandate cybersecurity controls as essential safety requirements, including encryption, authentication, and ongoing security updates. Services we identify as vulnerable (see \cref{tab:policies}) represent either: (i)~non-certified products deployed in violation of regulations, or (ii)~certified products operated outside their labeled security requirements. Both scenarios expose healthcare organizations to regulatory enforcement.



\paragraph{Data Protection \& Privacy}
Moreover, the technical vulnerabilities we measure directly violate data protection requirements under HIPAA~\cite{hipaa_164} (US), GDPR~\cite{gdpr_32} (EU), and equivalent frameworks worldwide. Specifically, Internet exposure without encryption breaches GDPR's confidentiality requirements (Art. 32) and HIPAA's transmission security standards (§ 164.312(e)). Similarly, the lack of authentication and authorization mechanisms violates GDPR's integrity and access control mandates (Art. 32) alongside HIPAA's access control specifications (§ 164.312(a)). More broadly, the inadequate security measures we observe fail both frameworks' fundamental requirements for implementing appropriate technical safeguards to protect patient data.

\section{DICOM Exposure}
\label{sec:active}

This section analyzes the results of our scanning campaign to characterize the security posture of Internet-facing \gls*{dicom} services.
We begin by describing the pre-processing pipeline used to remove false positives (noise such as honeypots and other measurement artifacts), then assess the security weaknesses present in the remaining real deployments.
Throughout, we evaluate the confidentiality, authentication, and authorization properties of exposed services.

\cref{tab:summary} summarizes our findings per port, reporting the number of responding addresses (Results), confirmed \gls*{dicom} services (Exposed), filtered false positives (Noise), and services meeting our vulnerability criteria (Vulnerable).
Total unique host counts are shown in the last row; note that a single host may expose multiple services.

\begin{table}
\centering
\caption{
Results from the \gls*{dicom} scanning campaign.
Observations from multiple iterations are coerced according to our pre-processing methodology.
Rows show the number of observations per port; columns are summarized at the bottom of the table, showing the number of unique hosts with one or more observations.
Scan results indicate the number of responding addresses.
Exposed counts observed deployments.
Noise counts filtered false-positives.
Vulnerable counts exposed services fulfilling our vulnerability criteria.
}
\label{tab:summary}
\begin{tblr}{
  width=\columnwidth,
  colspec={l | X X X X c},
  row{odd[2-Z]}  = {bg=gray!10},
  row{even[2-Z]} = {bg=white},
  row{1} = {font=\bfseries, halign=l},
  column{1} = {halign=l},
}
Port & Results & Exposed & Noise & Vulnerable\\
\midrule
104    & 2,390,337 & 3,666 & 2,085 & 1,574 \\
2761   & 2,577,728 & 1,473 & 1,364 & 109 \\
2762   & 2,569,319 & 1,471 & 1,359 & 112 \\
4242   & 2,445,424 & 2,381 & 1,363 & 1,001 \\
11112  & 2,500,040 & 4,145 & 2,524 & 1,605 \\
\midrule
\textbf{Unique}  & \textbf{3,127,526} & \textbf{6,646} & \textbf{2,625} & \textbf{3,979} \\
\bottomrule
\end{tblr}
\end{table}

\subsection{Data Processing}
\label{sec:active-processing}

Active measurement datasets inherently contain a mix of irrelevant responses (\ie hosts offering unrelated services, refusing connections, or no longer reachable) and false positives introduced by deception deployments such as honeypots and network telescopes~\cite{mladenov2025all,YABEN2026104911}.
Neither category represents real \gls*{dicom} exposure, and considering them inflates vulnerability estimates.
Our pre-processing pipeline addresses both problems in three stages: (i)~cleaning true negative responses, (ii)~filtering false positives, and (iii)~classifying remaining services against our threat model.

\subsubsection{Cleaning Strategy}
\label{sec:cleaning}

This stage removes all records unrelated to \gls*{dicom} communications, specifically observations from unreachable hosts and connections terminated prematurely by unrelated services.
However, \gls*{dicom} refusals (\ie A-ASSOCIATE-RJ responses) are retained as positive observations, since a well-formed rejection confirms the presence of a \gls*{dicom}-speaking service and may itself disclose implementation metadata (cf. \cref{table:taxonomy}).
Similarly, sessions in which the association succeeded but the subsequent C-ECHO timed out are counted as exposed, since completing the handshake alone constitutes a meaningful security boundary crossing.
Together, these cases contribute to the 6,646 addresses identified as exposing one or more \gls*{dicom} services (cf. \cref{tab:summary}).

\subsubsection{False-positive Filtering}
\label{sec:honeypot-filtering}

Our false-positive filtering method identifies sources of deception and malformed responses through five criteria, summarized in \cref{tab:noise}.

\begin{table}
\centering
\caption{Noise filtering criteria.}
\label{tab:noise}
\begin{tblr}{
 width=\columnwidth,
 colspec={X r},
 row{odd[2-Z]}  = {bg=gray!10},
 row{even[2-Z]} = {bg=white},
 row{1} = {font=\bfseries, halign=c},
 column{1} = {halign=l},
}
Criteria & Hosts \\
\midrule
Unrealistic number of exposed \gls*{dicom} services ($n>3$) & 1,419 \\
Missing User Info Implementation UID & 1,194 \\
Identical responses to the requested parameters (\ie AE titles, User Info, or PDU Type) & 35 \\
Unexpected PDU Type & 31 \\
Hardcoded values & 1 \\
\bottomrule
\end{tblr}
\end{table}

\begin{figure}
    \centering
    \includegraphics[width=\columnwidth]{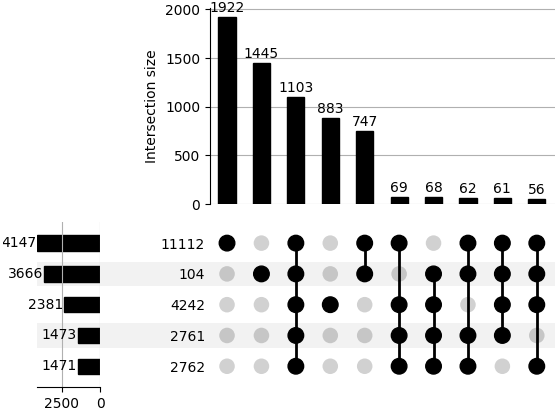}
    \caption{Distribution of apparent \gls*{dicom} exposure across scanned ports. Left bars show per-port counts, while upper bars show host intersections across port combinations. Large multi-port intersections highlight unrealistic deployment patterns, motivating our filtering of honeypots, telescopes, and other measurement artifacts.}
    \label{fig:ports}
\end{figure}

First, we follow the methodology of~\cite{YABEN2026104911,mladenov2025all}, which establishes that the probability of a host legitimately exposing multiple unrelated services simultaneously is low, and flag hosts that expose \gls*{dicom} on more than 3 ports simultaneously as likely telescopes or honeypots.
Across all 3,127,526 responding hosts, 50\% expose exactly one port and 95\% expose at most three. Among the 6,646 \gls*{dicom}-speaking hosts specifically, more than 65\% expose exactly one port and more than 76\% expose at most two, with the most common multi-port combination being ports 104+11112 (747 hosts), as illustrated in \cref{fig:ports}. By contrast, 1,103 hosts respond to all five scanned ports, a pattern inconsistent with real deployments.
We also validate this threshold against Censys~\cite{censys_2024_ioht}, which confirms that these hosts expose an abnormal number of unrelated services, and discard the 1,419 hosts exceeding it as likely telescopes.

Second, we flag 1,194 hosts (18\% of apparent \gls*{dicom} services) whose association responses omit the Implementation UID and/or Version Name fields required by the \gls*{dicom} standard~\cite{PS3729:online}.
Per the standard, association requests and responses must include a User Information object that specifies the Implementation UID (T=52H) and, optionally, a Version Name (T=55H).
This behavior, also present in the latest version of Dicompot, is consistent with fingerprint-evasion strategies and is absent from every \gls*{dicom} viewer we tested locally (cf. \cref{sec:exposure}).
In particular, none of the services showing this behavior matched the rest of the filtering criteria: they exposed at most two \gls*{dicom} services, with addresses scattered across the Internet.
We treat this as a honeypot heuristic while acknowledging it may not be unique to Dicompot.

Three additional criteria refine the filter: 35 hosts echo back identical parameters to the request, suggesting test servers or low-interaction honeypots; 31 hosts return PDU types not used for success or error messages; and one instance is identifiable directly from a hardcoded Dicompot Implementation UID, indicating an old version of the tool.
In total, our noise filtering criteria removes 2,625 addresses, reducing apparent exposure by 39\%.

\subsubsection{Vulnerability Identification}
\label{sec:vulnerability-identification}

After filtering, we evaluate the remaining 4,021 addresses against our threat model.
\cref{tab:policies} summarizes each evaluation policy, its threat level, and the number of matching hosts.
Addresses are evaluated against all applicable policies and assigned the highest matching threat level.

Level~1 indicates \gls*{dicom} exposure with some form of access control in place: 42 services responded to our probe but rejected the association.
Level~2 applies to services that completed the association handshake despite the absence of User Identity credentials in our probe; among these, 1,782 run outdated or deprecated software, and 85 are non-certified products identified in verifiably medical environments.
Level~3 covers services that also accept unauthenticated C-ECHO commands; 1,551 of these carry known remotely exploitable vulnerabilities.
We deliberately stop at Level~3 to avoid retrieving patient records (cf. \cref{table:taxonomy}).
Mitigations for identified weaknesses are discussed at the end of this section.

\begin{table}
\centering
\caption{Evaluation policies to apply our threat model.}
\label{tab:policies}
\begin{tblr}{
  width=\columnwidth,
  colspec={X c r},
  row{odd[2-Z]}  = {bg=gray!10},
  row{even[2-Z]} = {bg=white},
  row{1} = {font=\bfseries, halign=c},
  column{1} = {halign=l},
}
Policy & Level & Hosts \\
\midrule
Service identified, but association refused & 1 & 42 \\
Association accepted without providing authentication parameters & 2 & 3,979 \\
Outdated software & 2 & 1,782 \\
Non-certified software deployed in medical environments & 2 & 85 \\
Contains known vulnerabilities remotely exploitable & 3 & 1,551 \\
Accepts unauthenticated commands from unauthorized clients & 3 & 3,979 \\
\bottomrule
\end{tblr}
\end{table}
 
\subsection{Results}
\label{sec:exposure}

\subsubsection{Fingerprinting Services \& Information Verification}

\gls*{dicom} services are fingerprinted using the Implementation UID and Version Name fields returned during the association handshake.
In most cases, these fields provide sufficient detail to identify the vendor, product, and version of the exposed service.

The Implementation UID must be unique and registered to claim conformance with the \gls*{dicom} standard~\cite{CDICOMUn18:online}.
Although the standard imposes no fixed structure beyond an organizational root value, our results show that it is common practice to append product- and version-specific suffixes.
Where the UID alone is insufficient, vendor and version details can often be derived from the Implementation Version Name field.
No authoritative registry exists to serve as a matching baseline; services such as the OID Repository~\cite{OIDrepos72:online} do not maintain a complete listing of assigned root UIDs.

To address this, we (i)~grouped Implementation UID and Version values into a similarity tree and (ii)~developed individual parsing functions for all products observed more than once, yielding 21 clusters with recognizable naming patterns.
To confirm that these clusters represent real services rather than deception systems using spoofed identifiers, we manually inspected the source code of open-source implementations (OFFIS DCMTK, Pydicom, dcm4che, and Fellow OAK DICOM) and consulted \gls*{dicom} conformance statements, product manuals, and vendor documentation (including Fujifilm Synapse PACS, RadiAnt, and Santesoft, among others).
Further, we tested several products from 5 of the 15 identified vendors locally.
\cref{tab:products} lists the products identified and count number of observations, along with whether each was set up locally or was an open-source \gls*{dicom} library or toolkit.


\begin{table}
\centering
\caption{
List of identified \gls*{dicom} services.
The table includes vendor and product names, as well as root Implementation UIDs, and the number of observed deployments.
Open-source projects and locally deployed products are indicated.
}
\label{tab:products}
\begin{threeparttable}

\begin{tblr}{
  width=\columnwidth,
  colspec={X X X X[r]},
  cells = {font = \fontsize{7pt}{7pt}\selectfont},
  row{odd[2-Z]}  = {bg=gray!10},
  row{even[2-Z]} = {bg=white},
  row{1} = {font=\bfseries, halign=c},
  column{1} = {halign=l},
}

Vendor & Product & Root UID & Count \\
\midrule
OFFIS & DCMTK~\tnote{1} & 1.2.276.0.7230010.3 & 1,477 \\
dcm4che & dcm4che~\tnote{1} & 1.2.40.0.13.1.3 & 511 \\
Java DICOM Toolkit & jdt~\tnote{1} & 1.2.826.0.1.3680043.2.60.0.1 & 464 \\
Pydicom & Pynetdicom~\tnote{1} & 1.2.826.0.1.3680043.9.3811 & 102 \\
Fujifilm & Synapse PACS & 1.2.840.113845.1.1 & 70 \\
Fellow OAK DICOM & fo-dicom~\tnote{1} & 1.3.6.1.4.1.30071.8 & 65 \\
Neologica & LogiPACS & 1.3.6.1.4.1.18047 & 48 \\
CharruaSoft & CharruaPACS~\tnote{2} & 1.2.826.0.1.3680043.2.1396 & 34 \\
Asteris & Keystone-3D & 1.2.276.0.2783747.3.0.3 & 24 \\
RamSoft & powerserver & 1.2.124.113540.1 & 24 \\
NovaRad & NovaRad & 1.2.840.114051 & 20 \\
GE HealthCare & ViewPoint & 1.2.840.113619.6 & 14 \\
Microsoft & IIS & 1.2.826.0.1.3680043.2.360 & 13 \\
MedWeb & MedWeb & 2.16.840.1.113888.0.2.0.7 & 11 \\
Carestream & Image Suite & 1.2.840.113564.86.2 & 10 \\
RadiAnt & RadiAnt viewer~\tnote{2} & 1.2.826.0.1.3680043.8.1223.1.2 & 9 \\
Santesoft & Santesoft~\tnote{2} & 1.2.300.0.7238010.5 & 7 \\
Cybermed & OnDemand3D~\tnote{2} & 1.2.410.200034.0.0 & 5 \\
ONIS & ONIS~\tnote{2} & 1.2.392.200193 & 2 \\
\bottomrule

\end{tblr}

\begin{tablenotes}
  \item[1] Open-source toolkit and/or library.
  \item[2] Deployed and tested locally.
\end{tablenotes}

\end{threeparttable}
\end{table}

\subsubsection{Workstations \& Servers}

Identifiable exposed \gls*{dicom} services span \gls*{pacs} servers, viewer workstations, and worklist servers, though the majority rely on open-source \gls*{dicom} libraries embedded in printers, radiology devices, and custom servers.
These systems serve distinct roles: \gls*{pacs} servers act as centralized distribution hubs, while workstations handle images locally, and radiology devices transmit images on demand.
Because \gls*{dicom} images may reside on any of these systems, all are potential targets for data exfiltration.
It is worth noting that the \gls*{dicom} protocol does not support remote control of imaging hardware.

\textbf{Overview.}
Our dataset contains 3,979 exposed and vulnerable \gls*{dicom} systems, of which 2,910 could be attributed to known products (cf. \cref{tab:summary}).
The remaining 1,069 systems use open-source libraries in configurations that resist attribution; their clustering produced no similarity groups with more than two observations each.
Of the attributed services, 256 map to 11 distinct \gls*{pacs} products: Fujifilm Synapse leads with 70 instances, followed by Neologica LogiPACS with 48.
An additional 35 observations correspond to 3 \gls*{dicom} viewer products.
Open-source toolkits dominate the dataset; in some cases, AE titles and library version strings allow weak inference of the underlying application (\eg Horos~\cite{horospro17:online} and Slicer3D~\cite{SlicerSl79:online} viewers, whose latest releases are built on OFFIS DCMTK $v3.6.7$).

\begin{figure*}
    \centering
    \includegraphics[width=\textwidth]{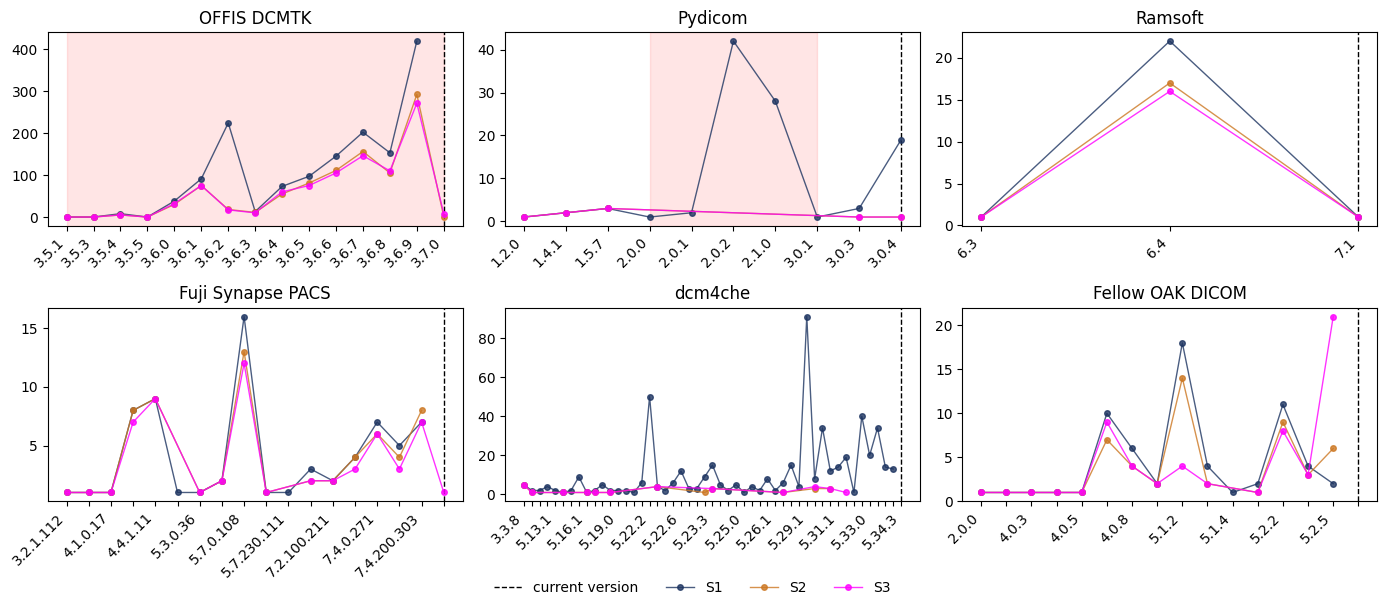}
    \caption{
    Version distribution of exposed services with the largest variance and the highest number of observations across scans.
    $S1$=January, $S2$=April, and $S3$=May (2026).
    The X-axis shows the version of each service as encountered; the Y-axis shows the observation count.
    Red zones indicate versions affected by remotely exploitable known vulnerabilities.
    The vertical dashed line marks the most recently released known version.
    }
    \label{fig:versions}
\end{figure*}

\textbf{Version and Software Distribution.}
\cref{fig:versions} shows the version distribution for the most prevalent products across all three scans, capturing more than 75\% of identified deployments.
Two patterns stand out.
Newer software versions are substantially more common, consistent with active maintenance in at least part of the deployment population.
At the same time, a persistent long tail of older versions remains, and 1,551 services run versions carrying known remotely exploitable vulnerabilities.

OFFIS DCMTK illustrates both dynamics.
We observe 420 instances on $v3.6.9$ (the latest available version at the time of $S1$), yet that version maintains CVE-2026-5663, a remotely exploitable arbitrary code execution flaw disclosed shortly after $S2$.
At the other end, 700 instances run versions below $v3.6.7$, which carry CVE-2022-2120, a comparable vulnerability from 2022.
The coexistence of recent and years old vulnerable deployments points to reactive rather than systematic patching behavior.
It is consistent with long-term stagnation observed in adjacent fields of \gls*{ot} and critical systems~\cite{7906943, 11129560}, in which deployments receive infrequent or no maintenance until retirement.

\begin{figure}
    \centering
    \includegraphics[width=\columnwidth]{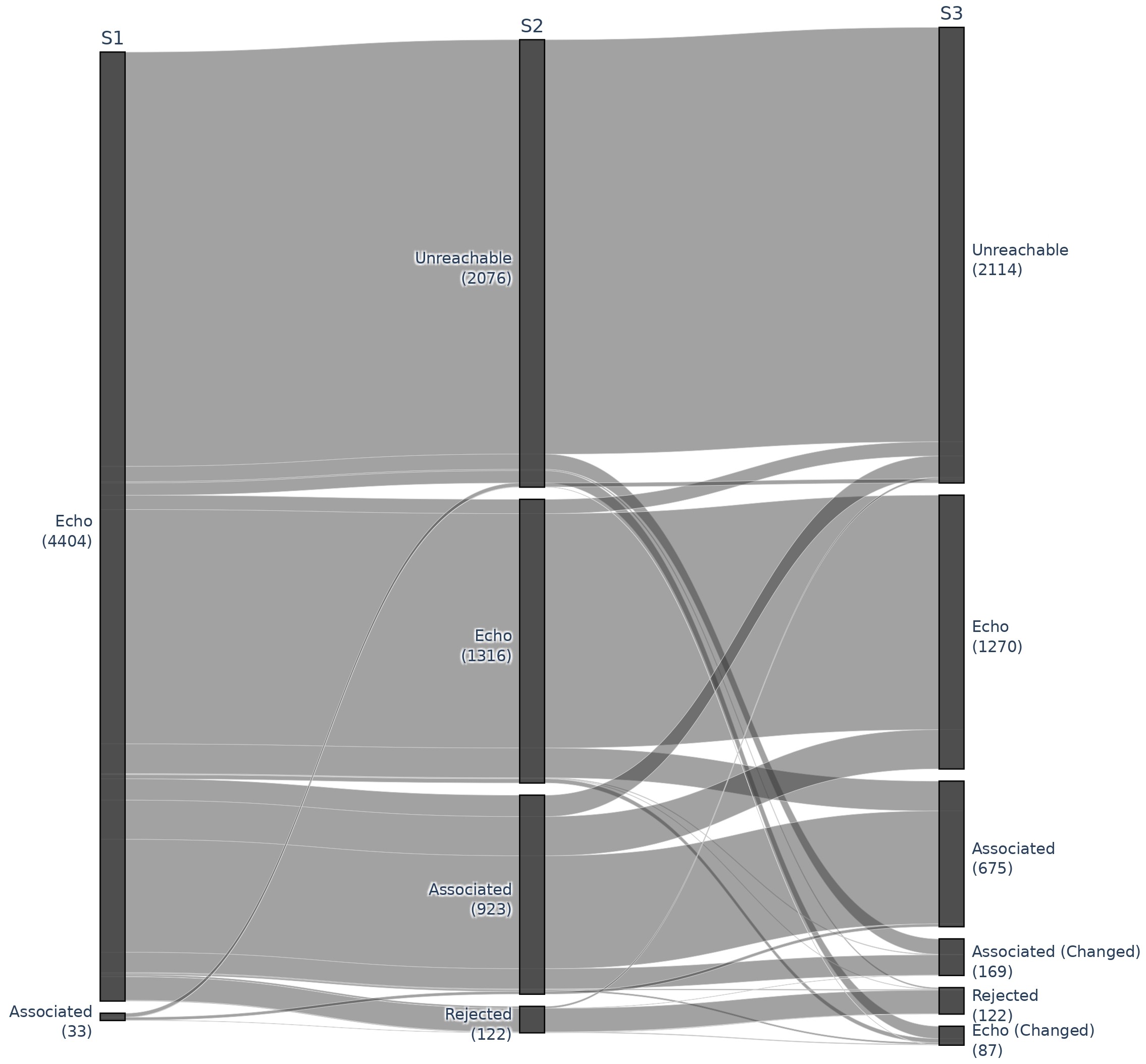}
    \caption{Parallel categories plot showing the evolution of host states across three observations (in order of sequence): $S1$, $S2$, and $S3$.
    Each line represents a host--port pair and traces its transition between states (\eg reachable, echoed, associated, unreachable) over time.
    Band width reflects the number of hosts following each path, highlighting common patterns of stability, change, and loss of connectivity.}
    \label{fig:changes}
\end{figure}

\textbf{Monitoring.}
To test this assessment, we conducted a second scan in April 2026, before our responsible disclosure campaign ($S2$), and a third in May 2026, following it ($S3$).
\cref{fig:changes} traces the state transitions of individual host-port pairs across all three observations.

Several findings emerge.
Approximately 50\% of services reachable in $S1$ were unreachable by $S2$, and fewer than 5\% of those subsequently recovered---all of which had received updates, suggesting that disappearances largely reflect transient deployments or address churn rather than deliberate remediation.
Of the services that remained reachable, 87 showed evidence of software updates between scans, the most common being OFFIS DCMTK upgrades to $v3.7$ following CVE-2026-5663.
This reactive behavior is encouraging but narrow: no analogous activity was observed among services running versions below $v3.6.7$, despite CVE-2022-2120 having been public since 2022.

Our responsible disclosure campaign produced only a modest shift in service behavior.
Fewer than 5\% of state transitions between $S2$ and $S3$ represent a meaningful reduction in threat level (\eg from accepting C-ECHO to rejecting associations or becoming unreachable), and we cannot attribute even this change specifically to our outreach rather than natural churn.
We therefore conservatively estimate that up to 50\% of exposed \gls*{dicom} services are effectively stagnant, receiving neither security maintenance nor any response to remediation efforts in the short term.

This pattern closely resembles that of other long-lifecycle \gls*{ot} deployments, where scheduled maintenance windows, hardware constraints, and vendor support limitations combine to make timely patching structurally difficult~\cite{stouffer2011guide, krotofil2013industrial}.

\textbf{Compliance.}
The security weaknesses we document carry direct regulatory consequences under both HIPAA (§\,164.312) and GDPR (Art.~32), which require encryption in transit, access controls, and audit mechanisms for systems handling protected health information.
None of the services we examined provided encryption by default, and none enforced authentication at the \gls*{dicom} protocol level.

A plausible but ultimately insufficient explanation is that most exposed systems operate outside clinical contexts (\eg in research, veterinary, or laboratory settings) and therefore contain only anonymized or test data.
The findings of \citet{apilite2023} reject this interpretation: they show that 3,806 exposed \gls*{dicom} services collectively exposed over 16 million personal records accessible without authentication, a figure closely resembling the number of vulnerable services in our dataset.
This motivates two alternative explanations: (i)~clinics and medical institutions may be breaching compliance by deploying non-certified products that do not meet applicable security standards; or (ii)~\gls*{dicom} services may be unintentionally exposed due to network or device misconfiguration, for instance when \gls*{pacs} systems that offer patient portals are not shielded by appropriate firewall rules.
Neither explanation justifies the absence of encryption, authentication, and access control; weaknesses that enable unauthorized clients to retrieve, modify, upload, or delete \gls*{dicom} documents regardless of deployment context.

\subsubsection{Verification}
To investigate the presence of medical institutions in our dataset, we enrich our results with organizational metadata from IPinfo, correlating IP addresses with hospital and clinic identifiers, domain names, and WHOIS records.
Attribution remains an open challenge in Internet-wide measurement, and IPinfo does not guarantee complete coverage; we treat this enrichment as a first step rather than a definitive classification.

Following this process, we identified five addresses associated with hospitals and several additional medical clinics, which were verified manually through web portals and WHOIS cross-checks.
Notably, all institutions we could identify exposed a web portal, a detail that future measurement studies targeting medical infrastructure should incorporate to improve attribution accuracy.

Of the five hospitals, only one remained reachable through $S3$: a European hospital running OFFIS DCMTK $v3.6.6$, a severely vulnerable version.
The remaining four exposed dated versions of dcm4che ($v5.23.2$ to $v5.29.2$).
Manual inspection of clinic-associated records further identified deployments of non-certified products, including CharruaSoft and SanteSoft \gls*{pacs} servers.
By contrast, portals associated with certified \gls*{pacs} systems, most notably Neologica's LogiPACS, consistently refused our requests with HTTP 403 responses, suggesting that application-layer authorization controls are in place even where the underlying \gls*{dicom} service remains Internet-accessible and responds to C-ECHO.
This distinction does not eliminate risk, but it does indicate a degree of defense-in-depth absent from non-certified deployments.

Together, these observations establish that the \gls*{dicom} exposure problem is not confined to a single product class, certification status, or deployment context.
They also support the hypothesis that medical institutions are using non-certified products for diagnostic purposes.

\subsubsection{Local Deployments}
\label{sec:local-deployments}

To understand the root causes of \gls*{dicom} exposure and the security weaknesses we observe, we requested software from the identified vendors and deployed it locally.
Our testing targets two goals: (i)~identifying behaviors that diverge from what we observed in the wild, and (ii)~characterizing the installation process for security pitfalls and unsafe default configurations.

We evaluate each deployment against the security requirements defined by the \gls*{fda} and \gls*{ce} marking, using our threat model as a baseline to assess encryption, authentication, and authorization settings.
Each product is tested under both its default configuration and its most permissive allowed configuration.
The experimental environment is a Windows~11 Professional virtual machine with the latest security updates installed and the firewall turned off to allow inbound connections.
After installation, we insert a dummy \gls*{dicom} file from the Cancer Imaging Archive (TCIA)~\cite{WelcomeTCIA} and probe each service with our ZGrab2 scanner to assess access granted to unauthenticated clients.

In total, we tested four \gls*{pacs} servers and two \gls*{dicom} viewers from five of the fifteen identified vendors; software samples and test results are included in the submitted artifacts.
Two products, CharruaPACS and OnDemand3D, expose administrative panels protected only by default credentials, granting access to all server resources in our test environment.
Unencrypted \gls*{dicom} servers start by default in the majority of cases, including both standard ports 104 and 11112; encrypted communications are not enabled by default, likely due to the requirement for a valid certificate.
The RadiAnt viewer is the only product requiring manual input to start the \gls*{dicom} server.
Santesoft restricts inbound connections to localhost by default and uses non-standard ports 11121 and 11122, partial mitigations that alone do not constitute adequate security controls.
Only ONIS enables AE title verification and IP safelisting by default.
No tested product enforces authentication for \gls*{dicom} associations, and no product encrypts \gls*{dicom} files at rest or in transit.
Across all base installations, any client completing the association handshake is permitted to execute the full range of \gls*{dimse} operations.

\textbf{Takeaway.}
Reaching a vulnerable state requires multiple steps: Windows firewall rules must be disabled, and the \gls*{dicom} server must be configured to accept remote connections and permit all \gls*{dimse} operations.
Most servers initiate this process automatically and do not hide configuration options, though some require manual intervention to turn off validation features.
The fundamental obstacle, however, is structural. 
The absence of encryption and authentication as \emph{default} behaviors across \gls*{dicom} software means that exposure is a misconfiguration away for almost any deployment.

\subsubsection{Recommendations for Mitigation}

We propose the following recommendations in order of priority.

\textbf{Remove \gls*{dicom} Services From Internet Exposure.}
No service we identified has an operational requirement for direct Internet reachability.
Most \gls*{pacs} systems already provide web and WADO interfaces that offer equivalent capabilities with authenticated browser access.
Where remote \gls*{dicom} connectivity is strictly necessary, access should be gated through a VPN rather than exposed directly.

\textbf{Enforce Encryption, Authentication, \& Authorization.}
The \gls*{dicom} standard already supports TLS, User Identity negotiation, and AE title filtering.
The obstacle is not capability but the default configuration.
AE title validation and non-standard ports offer negligible protection against basic enumeration and should not substitute for proper authentication.

\textbf{Encrypt \gls*{dicom} Data at Rest.}
Regulatory frameworks in both the US and EU require protection of health records at rest, yet none of the services we tested encrypted stored files.
Vendors and operators should treat this as a baseline requirement.
The findings of \cite{apilite2023} reinforce this: patient-identifiable information was directly retrievable from exposed services, indicating that encryption at rest is not common practice.

\textbf{Establish Proactive Patching Schedules.}
Two new remotely exploitable vulnerabilities emerged between $S1$ and $S3$, and both triggered only limited reactive patching.
The long tail of unpatched deployments (many of which show no evidence of maintenance over years) reflects a structural challenge common to long-lifecycle \gls*{ot} devices, as well as a lack of urgency.
We recommend scheduled maintenance cycles and continuous monitoring of exposed services as basic security hygiene, rather than reactive patching triggered only by public CVE disclosure.
\section{Passive Measurements with Dicompot}
\label{sec:dicompot}
To complement our active scanning, we conducted passive Internet measurements by deploying Dicompot, the primary open-source \gls*{dicom} honeypot, to observe real-world exploitation attempts against exposed \gls*{dicom} services. 
This section reports the attacks we observed, the methodology we developed to isolate them from background noise, and the architectural limitations of the tool that bound our conclusions. Unlike our active campaign, which measures exposure at scale, this passive component is designed to characterize the typology of attacks targeting exposed \gls*{dicom} services rather than quantify their volume.

\subsection{Setup}
We deployed two instances of Dicompot: one within our university network environment, and another on a cloud instance located in Frankfurt, Germany. Both instances exposed default \gls*{dicom} ports 104 and 11112 and were deployed via the T-Pot framework. Data was collected over a 324-day period, from June 7, 2025, to April 26, 2026. 
While the extended timeframe ensures sufficient exposure to opportunistic attackers, we acknowledge inherent limitations in these vantage points. Specifically, due to operational constraints, we could not deploy the honeypots within a verified medical facility's IP space. 
To approximate a realistic threat landscape, our university network served as the most viable alternative, as academic networks frequently host legitimate, research-oriented medical imaging infrastructure, providing a plausible decoy environment for attackers. Furthermore, the cloud instance in Frankfurt provided a secondary geographic vantage point, although commercial cloud subnets are less commonly associated with on-premises hospital infrastructure.
Despite these constraints, our dual deployment provides sufficient visibility to establish a reliable baseline of opportunistic attacks, complementing the findings of our active measurement campaign.

\subsection{Results}
\label{sec:dicompot_results}

\begin{figure*}
    \centering
    \includegraphics[width=\textwidth]{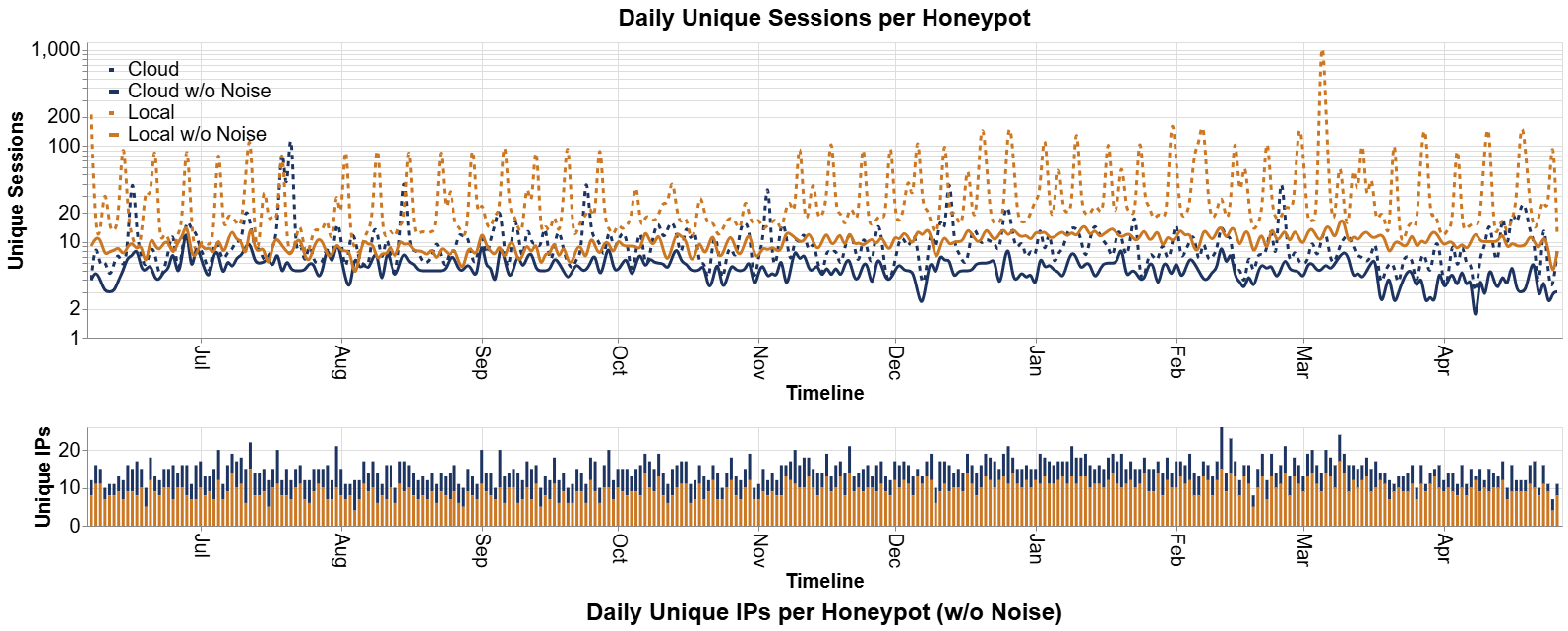}
    \caption{Comparison Timeline of Sessions with the \gls*{dicom} Service, highlighting the impact of filtering non-\gls*{dicom} noise.}
    \label{fig:dicom_timeline}
\end{figure*}

\textbf{Data Sanitization.}
Before reporting attack observations, we must address a critical artifact in Dicompot's raw logs: Its logging mechanism is highly prone to recording false-positive interactions. As shown in \cref{fig:dicom_timeline}, Dicompot recorded substantial and regular traffic spikes. However, cross-referencing IP addresses and conducting local \texttt{nmap} testing revealed that Dicompot logs generic TCP connection attempts (\eg SYN scans, malformed packets) as legitimate \gls*{dicom} sessions, heavily inflating the perceived threat level. The regular spikes observed on the local deployment were actually caused by an internal, non-\gls*{dicom} specific scanner operated by our university.
To isolate high-confidence \gls*{dicom} interactions, we developed a strict filtering heuristic: we discarded any session lacking a client \texttt{Implementation Version Name}. Although technically defined as an optional field in the \gls*{dicom} standard, it is universally transmitted by legitimate \gls*{dicom} clients in practice and entirely absent in generic TCP noise. We derived this filtering approach through controlled local testing; while it serves as an empirical approximation, it is consistent with our \texttt{nmap} ground truth and provides a robust baseline for isolating genuine \gls{dicom} interactions.
As illustrated in \cref{fig:dicom_timeline} and summarized in \cref{table:hp_stats}, applying this filter reduced total sessions from 23,405 to 4,998, with the local deployment seeing an 83\% reduction (19,358 to 3,242) and the cloud deployment a 57\% reduction (4,047 to 1,756).

\begin{table}
    \centering
    \caption{General statistics for sessions across both Dicompot deployments, illustrating the impact of noise filtering. \gls*{dicom}-specific commands are extracted from the filtered high-confidence dataset.}
    \label{table:hp_stats}
    \begin{tabular}{@{}l S[table-format=5.0] S[table-format=4.0] S[table-format=4.0] S[table-format=4.0]@{}}
        \toprule
        & \multicolumn{2}{c}{\textbf{Local Environment}} & \multicolumn{2}{c}{\textbf{Cloud Environment}} \\
        \cmidrule(lr){2-3} \cmidrule(l){4-5}
        \textbf{Metric} & {\textbf{Raw}} & {\textbf{Filtered}} & {\textbf{Raw}} & {\textbf{Filtered}} \\
        \midrule
        Sessions    & 19358 & 3242 & 4047 & 1756 \\
        Unique IPs  & 1993  & 1201 & 1459 & 974  \\
        \midrule
        C-ECHO      & {--}  & 96   & {--}  & 116  \\
        C-FIND      & {--}  & 7    & {--}  & 8    \\
        C-GET       & {--}  & 0    & {--}  & 0    \\
        \bottomrule
    \end{tabular}
\end{table}

\textbf{Attacks.}
Having isolated high-confidence \gls*{dicom} traffic, we identify 15 C-FIND commands that we confidently classify as deliberate data exfiltration attempts. While the majority (13) targeted the \texttt{PatientName} attribute with a global wildcard (\texttt{*}), one actor issued two commands with queries containing no search filters, effectively requesting the entire database. In all cases, these commands successfully retrieved all 50 fake patient records stored in the honeypots. These query patterns are strong indicators of malicious intent: Legitimate clinical users search for specific patient identifiers, not unrestricted dumps of an entire archive. These attempts originated from three distinct sources, detailed in \cref{tab:attacks}, each exhibiting different tooling.

\begin{table}
    \centering
    \caption{Summary of representative C-FIND attacks observed on Dicompot.}
    \label{tab:attacks}
    \begin{tabular}{@{}llll@{}}
        \toprule
        \textbf{Category}  & \textbf{Client} & \textbf{CMD} & \textbf{Search} \\
        \midrule
        Academic Scan.  & \texttt{PYNETDICOM\_310} & C-FIND & * \texttt{PatientName} \\
        Private Network   & \texttt{RadiAnt-2025.1}  & C-FIND & \textit{(no filter)} \\
        Hosting Provider & \texttt{OFFIS\_DCMTK\_369} & C-FIND & * \texttt{PatientName} \\
        \bottomrule
    \end{tabular}
\end{table}

The most active attacker was responsible for 12 of the 15 C-FIND commands, issuing 6 queries against each deployment across three separate days, with each day's activity concentrated within a roughly one-hour window; a pattern consistent with systematic, automated harvesting rather than opportunistic probing.
This actor used the \texttt{PYNETDICOM\_310} library and accounted for all honeypot observations of \texttt{PYNETDICOM\_210} and \texttt{GODICOM\_1\_1}, suggesting a single operator rotating tooling across sessions. Notably, this IP belongs to an academic institution whose public-facing webpage explicitly states it conducts only non-malicious scans. Yet, since global C-FIND queries actively extract potentially sensitive patient data, deploying them in internet-wide scans raises significant ethical concerns. This stands in direct contrast to our own methodology (\cref{sec:methods}), which deliberately excludes all data retrieval commands.
The second actor, originating from a private German network, issued two C-FIND commands exclusively against the cloud instance using \texttt{RadiAnt-2025.1}, a commercial \gls*{dicom} viewer not typically associated with automated scanning, suggesting a more targeted, manual interaction. Unlike the other attackers, this actor issued C-FIND commands without specifying a search attribute, yet still retrieved all stored records by relying on the server's default behavior.
Finally, the third source, from a commercial hosting provider, issued a single C-FIND against the local deployment using \texttt{OFFIS\_DCMTK\_369}.

Beyond C-FIND activity, we observed 212 C-ECHO requests originating from 56 and 75 unique IPs on the local and cloud deployments, respectively (\cref{table:hp_stats}), consistent with automated connectivity verification preceding potential exploitation (a reconnaissance pattern mirroring the C-ECHO-then-C-FIND sequence documented in our threat model in \cref{sec:threat-model}). The fact that most of these IPs never escalated beyond C-ECHO suggests that a non-trivial fraction of actors may have recognized the decoy nature of the honeypot and abandoned their sessions before proceeding, implying that our 15 observed C-FIND attacks likely underestimate the true number of actors with data-harvesting intent. Moreover, no C-GET commands were observed, likely reflecting Dicompot's limited interactivity causing advanced actors to abandon the session before escalating beyond reconnaissance, a point we return to below.

The software distribution in \cref{table:version} is also informative: \texttt{OFFIS\_DCMTK} variants dominated traffic at nearly 98\% of filtered sessions. However, since software versions can be easily spoofed within the \gls*{dicom} protocol, these values serve as behavioral indicators rather than definitive attribution. As expected, in the sanitized dataset, we also observed and verified traffic originating from our own \texttt{ZGRAB2} active scanner.

\begin{table}
    \centering
    \caption{Software used to access the local and cloud honeypots.}
    \label{table:version}
    \begin{tabular}{@{} l rrrr @{}}
        \toprule
        \textbf{Software} & \multicolumn{2}{l}{\textbf{Local}} & \multicolumn{2}{l}{\textbf{Cloud}} \\
        \midrule
        OFFIS\_DCMTK\_362 & 1,808 & (55.77\%) & 1,600 & (91.12\%) \\
        OFFIS\_DCMTK\_360 & 1,379 & (42.54\%) & 95    & (5.41\%) \\
        CharruaVista      & 17    & (0.52\%)  & 22    & (1.25\%) \\
        PYNETDICOM\_210   & 13    & (0.40\%)  & 13    & (0.74\%) \\
        GODICOM\_1\_1     & 10    & (0.31\%)  & 10    & (0.57\%) \\
        ZGRAB2            & 8     & (0.25\%)  & 8     & (0.46\%) \\
        PYNETDICOM\_310   & 6     & (0.19\%)  & 6     & (0.34\%) \\
        RadiAnt-2025.1    & 0     & (0.0\%)   & 2     & (0.11\%) \\
        OFFIS\_DCMTK\_369 & 1     & (0.03\%)  & 0     & (0.0\%) \\
        \bottomrule  \end{tabular}
\end{table}

\textbf{Dicompot Limitations.} 
The attack counts reported above should be interpreted as a lower bound on real-world \gls*{dicom} exploitation activity. As discussed above, the gap between the 131 unique IPs issuing C-ECHO requests and the 3 actors escalating to C-FIND is potentially explained by Dicompot's detectability: Shodan successfully fingerprinted both instances as honeypots, and our own active measurement pipeline independently flagged both IPs as deception noise (\cref{sec:methods}). Sophisticated actors probing for complex protocol behaviors are likely to abandon the session upon identification, leaving our results primarily representative of opportunistic, automated activity. This is further compounded by the fact that the underlying Dicompot repository is no longer actively maintained, limiting its protocol fidelity and its utility as a long-term measurement platform. In particular, Censys did not apply deception tags to either instance, suggesting that detectability varies across passive intelligence platforms. 
Ultimately, the limited engagement from advanced adversaries underscores a significant gap in current medical deception systems. This is a challenge that future honeypot designs must address by implementing deeper protocol emulation and more realistic data environments to capture the full spectrum of targeted healthcare threats.

\section{Conclusion}
\label{sec:concl}
This paper presented a comprehensive, noise-aware evaluation of the global \gls*{dicom} security landscape, uniquely combining ethics-constrained active Internet-wide scanning with passive honeypot observations.

\textit{Active Measurements.}
We conducted active Internet-wide surveys to detect \gls*{dicom} exposure and evaluate the evolution of their security weaknesses over a period of five months on three separate occasions: January ($S1$), April ($S2$), and May ($S3$).
Our results show a concerning general absence of encryption, authentication, or authorization controls across 3,979 systems, regardless of software, certifications, or environment, with 1,551 of these affected by known remotely exploitable vulnerabilities.
While our findings suggest a moderate presence of maintenance, with nearly 50\% of services becoming unreachable shortly after our first scan, and numerous instances receiving updates after our responsible disclosure campaign (mostly attributed to security vulnerabilities released between $S1$ and $S2$), the systemic lack of default security configurations remains a critical threat to healthcare infrastructure.

\textit{Passive Measurement.}
On the passive measurement side, our 324-day Dicompot deployment highlights a fundamental limitation in current medical honeypot systems. After filtering generic TCP noise (which inflated raw session counts by 57–83\% depending on the location), we observed 15 C-FIND data-exfiltration attempts across both vantage points. The disproportionately low ratio of escalated attacks to initial C-ECHO probes strongly suggests that sophisticated actors identified the honeypot and disengaged. This conclusion is supported by Shodan's successful fingerprinting of both instances and our own active-measurement pipeline's independent detection.


\textit{Responsible Disclosure.}
The responses to our responsible disclosure campaign were mixed.
During this process, we asked vendors to verify our observations and requested software samples to continue our evaluation.
The majority of vendors responded to our messages with resistance or skepticism, blaming consumers for misconfigurations, or refusing to acknowledge security weaknesses. 
One of the vendors actively ignored our messages, categorizing the ticket we created as spam. 
However, we also learned about 147 honeypots mimicking the OsiriX software, deprecated services relying on old versions of libraries seen in the wild.
Lastly, we shared our findings and mitigation strategies with our regional \gls*{cert}, which manages the broader notification process to reach the rest of \glspl*{cert} and notify maintainers of the affected services.

Together, these findings establish that DICOM exposure is both widespread and structurally resistant to improvement under current practices. The vulnerabilities we document are not artifacts of a single vendor, product class, or deployment context: They reflect systemic failures of default configuration, maintenance discipline, and regulatory enforcement across the healthcare sector. Addressing them will require coordinated action from vendors, operators, and regulators: stronger security defaults, mandatory patching schedules, and clearer guidance on what Internet-facing DICOM deployments are and are not permissible under applicable data protection frameworks.

\textit{Future Work.}
Moving forward, our findings highlight two critical directions for future research. First, Dicompot's limited protocol fidelity and lack of active maintenance make it unsuitable as a long-term measurement platform: Future medical deception systems must implement deeper protocol emulation and realistic data environments to capture healthcare-targeted adversary behavior efficiently. Second, expanding the active measurement scope to encompass non-standard ports and auxiliary web portals will provide a more exhaustive assessment of the clinical attack surface, ultimately informing stronger default security standards for medical imaging protocols.


\newglossaryentry{ics}
{
    name={ICS},
    plural={ICSs},
    description={Industrial Control System},
    descriptionplural={Industrial Control Systems},
    first={\glsentrydesc{ics} (\glsentrytext{ics})},
    firstplural={\glsentrydescplural{ics} (\glsentryplural{ics})}
}

\newglossaryentry{ot}
{
    name={OT},
    description={Operational Technology},
    first={\glsentrydesc{ot} (\glsentrytext{ot})}
}

\newglossaryentry{cve}
{
    name={CVE},
    description={Common Vulnerabilities and Exposures},
    first={\glsentrydesc{cve} (\glsentrytext{cve})},
    plural={CVEs},
    descriptionplural={Common Vulnerabilities and Exposures},
    firstplural={\glsentrydescplural{cve} (\glsentryplural{cve})}
}


\newglossaryentry{dicom}
{
name={DICOM},
description={Digital Imaging and Communications in Medicine},
first={ \glsentrytext{dicom} (\glsentrydesc{dicom})},
}

\newglossaryentry{tcia}
{
name={TCIA},
description={The Cancer Imaging Archive},
first={\glsentrydesc{tcia} (\glsentrytext{tcia})},
}

\newglossaryentry{pacs}
{
name={PACS},
description={Picture Archiving and Communication System},
first={\glsentrydesc{pacs} (\glsentrytext{pacs})},
plural={PACSs},
descriptionplural={Picture Archiving and Communication Systems},
firstplural={\glsentrydescplural{pacs} (\glsentryplural{pacs})}
}

\newglossaryentry{sop}
{
name={SOP},
description={Service-Object Pair},
first={\glsentrydesc{sop} (\glsentrytext{sop})},
}

\newglossaryentry{dimse}
{
name={DIMSE},
description={DICOM Message Service Element},
first={\glsentrydesc{dimse} (\glsentrytext{dimse})},
plural={DIMSEs},
descriptionplural={DICOM Message Service Elements},
firstplural={\glsentrydescplural{dimse} (\glsentryplural{dimse})},
}

\newglossaryentry{iod}
{
name={IOD},
description={Information Object Definition},
first={\glsentrydesc{iod} (\glsentrytext{iod})},
}

\newglossaryentry{iana}
{
name={IANA},
description={Internet Assigned Numbers Authority},
first={\glsentrydesc{iana} (\glsentrytext{iana})},
}

\newglossaryentry{rce}
{
name={RCE},
description={Remote Code Execution},
first={\glsentrydesc{rce} (\glsentrytext{rce})},
}

\newglossaryentry{as}
{
    name={AS},
    description={Autonomous System},
    first={\glsentrydesc{as} (\glsentrytext{as})}
}

\newglossaryentry{mri}
{
name={MRI},
description={Magnetic Resonance Imaging},
first={\glsentrydesc{mri} (\glsentrytext{mri})},
}

\newglossaryentry{ct}
{
name={CT},
description={Computed Tomography},
first={\glsentrydesc{ct} (\glsentrytext{ct})},
}

\newglossaryentry{iomt}
{
name={IoMT},
description={Internet of Medical Things},
first={\glsentrydesc{iomt} (\glsentrytext{iomt})},
}

\newglossaryentry{ce}
{
name={CE},
description={Conformité Européenne},
first={\glsentrydesc{ce} (\glsentrytext{ce})},
}

\newglossaryentry{fda}
{
name={FDA},
description={Food and Drug Administration},
first={\glsentrydesc{fda} (\glsentrytext{fda})},
}

\newglossaryentry{cert}{
name={CERT},
description={Cyber Emergency Response Team},
first={\glsentrydesc{cert} (\glsentrytext{cert})},
plural={CERTs},
descriptionplural={Cyber Emergency Response Teams},
firstplural={\glsentrydescplural{cert} (\glsentryplural{cert})},
}

\newglossaryentry{pii}
{
name={PII},
description={Personally Identifiable Information},
first={\glsentrydesc{pii} (\glsentrytext{pii})},
}




\cleardoublepage
\appendices

\section{Ethical Considerations}
\label{sec:ethics}

In conducting this research, we adhered to the principles outlined in the Menlo Report. We recognize that active scanning of medical infrastructure and the usage of deception technologies involve distinct ethical risks. Below, we detail our stakeholder-based analysis and the mitigation employed to minimize harm.

\subsection{Stakeholder Analysis}
We identified three primary stakeholder groups impacted by our work: (1) Medical Institutions and System Owners (hospitals/clinics), (2) Patients (whose data resides on these systems), and (3) The Defensive Security Community (researchers using Dicompot).

\subsubsection{Impacts and Harms}
\begin{itemize}
    \item \textbf{Medical Institutions (System Availability)}. The primary risk to system owners during the measurement phase (\cref{sec:active}) was the potential for service disruption. Legacy medical devices often possess fragile network stacks; high-rate scanning could theoretically cause denial-of-service conditions or crash critical equipment.
    \item \textbf{Patients (Privacy)}. Although our scan sought to identify vulnerabilities, the \gls*{dicom} protocol transmits Patient Health Information (PHI). A significant potential harm would be the unauthorized collection, storage, or viewing of sensitive patient data (images or metadata) during the vulnerability assessment.
    \item \textbf{The Research Community (Efficacy of Existing Tools)}. By publishing the specific fingerprinting signatures of Dicompot (\cref{sec:exposure}), we may reduce the utility of existing Dicompot deployments. There is a risk that malicious actors could use our findings to bypass existing detection grids. 
    \item \textbf{The Research Community (Data Sharing)} Our collection of Internet scans also contains \gls*{pii} in the form of IP addresses. This information is necessary for the analysis, reproducibility, and comparison for future work.
    Similarly, our dataset of Dicompot logs contains IP addresses. Since IP addresses constitute \gls*{pii} under applicable data protection frameworks, they cannot be shared openly.
\end{itemize} 

\subsubsection{Mitigation}
\begin{itemize}
    \item \textbf{Scanning Hygiene}. To mitigate risks to medical institutions, we followed standard scanning best practices (ZMap/Masscan guidelines): Scans were performed at a low packet rate to prevent network congestion or service degradation (Rate Limiting); We strictly limited our interaction to the \gls*{dicom} Association negotiation (handshake) and basic C-ECHO verification. We did not issue other commands that manipulate imaging data (Non-Intrusive Probing);
    We respected all standard exclusion lists (\eg valid blacklist entries) and hosted a web page at the scanning IP address, providing opt-out instructions and researcher contact information. No complaints were received during the study (Exclusion Lists).
    
    \item \textbf{Privacy Preservation}. We operated under a strict \textit{Metadata Only} policy. Our probes were designed to discard any payload containing PHI immediately. No patient names, birthdates, or medical images were stored. We recorded only technical metadata (AE titles, implementation versions, error codes) necessary to distinguish valid \gls*{dicom} services from honeypots. 

    We decided to share our artifacts in a restricted manner through Zenodo (\cref{sec:open_science}). This allows us to retain necessary information, while ensuring responsible handling of the sensitive IP addresses.
    
    Regarding our passive measurement dataset, we pseudo-anonymized the IPs by replacing them with generic IDs, maintaining the relationships between the entries intact while removing the sensitive \gls*{pii} that is no necessary for reproducibility and further research.
    
    \item \textbf{Responsible Disclosure}. 
    For the results of our active measurements, we share our findings with the affected vendors and our regional \gls*{cert}.
    \gls*{cert} shared findings and mitigation strategies with affected system maintainers anonymously.
    
    Regarding the fingerprinting of Dicompot, we weighed the harm of exposing the flaws against the benefit of advancing the state of the art. Since the flaws identified are straightforward to detect (static responses), we determined that sophisticated attackers likely already possess this capability, as seen in \cref{sec:dicompot_results}. Therefore, hiding these flaws provides a false sense of security.
\end{itemize}

\subsubsection{Decision to Proceed}
We weighed the ethical harms against the benefits. The "Do Nothing" approach preserves the status quo, where medical institutions are unknowingly exposed, and defenders rely on easily detectable honeypots (Dicompot) that skew risk perception.
We concluded that the benefits, providing the first noise-free census of \gls*{dicom} exposure, 
outweigh the minimal risks associated with responsible scanning and the publication of honeypot signatures. Our decision prioritizes the Beneficence principle: Accurate measurement 
is a prerequisite for securing the Internet of Medical Things (IoMT).

\cleardoublepage

\section{Open Science} 
\label{sec:open_science}




To adhere to the open science policy, this appendix describes all artifacts necessary to evaluate and reproduce the scientific research of this paper, their availability, and restrictions arising from ethical or privacy concerns. All artifacts will be archived as described below in the persistent public repository Zenodo with the stable identifier (DOI) [ANONYMIZED] to ensure long-term availability and restrict access as described in Ethical Considerations (\cref{sec:ethics}). \textit{Note: To avoid exposure of reviewers, we temporarily make the data available in an anonymous repository\footnote{\url{https://anonymous.4open.science/r/DICOM_Internet_Measurement_Paper-8A55}} during the review process.}

\subsection{Metadata and Documentation}
We provide accompanying documentation and metadata for each artifact as necessary. This includes the structure of the dataset, data collection periods, processing steps applied to the raw data and requirements for running the code.

\subsection{Active Measurement}

\textbf{Dataset.}
We release a dataset with raw data collected during our Internet scans.
This includes ZMap and ZGrab2 results.
ZGrab2 results contain \gls*{pii}, specifically sensitive identifiers (IPs)---reason for selecting restricted sharing of our data.
However, we do not share any third-party data set such as the IPinfo dataset (\cref{sec:methods}), as this data is available through IPInfo.

\textbf{Source Code.}
A detailed methodology for scanning and data analysis is described in \cref{sec:methods,sec:active}. Following these methodologies, the raw dataset can be understood and processed. 
Additionally, we include the \gls*{dicom} probe such that researchers can conduct further scans.

\textbf{Local deployment artifacts.}
The binaries and samples deployed locally, along with summaries and other collected information from vendors.
In addition to files and templates used for the responsible disclosure campaign.

\subsection{Passive Measurement}
Our passive measurement (\cref{sec:dicompot}) contains artifacts for two aspects: first, we share the analysis code and data from our long-term Dicompot deployment; second, we provide the setup and results from the test of the heuristic for our data sanitization (\cref{sec:dicompot_results}).

\textbf{Datasets.}
We release a pseudo-anonymized dataset of our Dicompot deployments. The logs from both deployments are combined, and a field is added to indicate cloud and local deployment. The IPs are replaced with a generic label with a number as identifier (`Attacker<num>'). This allows us to remove the sensitive IP information while keeping enough information to redo our analysis.
Furthermore, we share the Dicompot logs and a PCAP file collected during our local testing of the heuristic. Since this was done on localhost inside a virtual machine, no further sanitization was necessary.

\textbf{Source Code.}
We share the Jupyter notebooks for the Dicompot log analysis and the Python script for the PCAP analysis, which can be used on the previously mentioned datasets to reproduce our findings.

\cleardoublepage


%
\bibliographystyle{IEEEtranN}
\bibliography{references}

\end{document}